\documentclass[reprint,amsmath,amssymb,aps]{revtex4-1}
\usepackage{graphicx} 
\usepackage{datetime}
\usepackage{dcolumn} 
\usepackage{bm} 
\usepackage{xcolor}

\DeclareMathOperator\arccosh{arccosh}

\begin{document}
\title{Surface-assisted carrier excitation in plasmonic nanostructures}
\author{Tigran V. Shahbazyan}
\affiliation{
Department of Physics, Jackson State University, Jackson, MS 39217 USA
}

\begin{abstract}
We present a quantum-mechanical model for surface-assisted carrier excitation by optical fields in plasmonic nanostructures of arbitrary shape. We derive an explicit expression, in terms of  local fields inside the metal structure, for surface absorbed power and surface scattering rate that determine  the  enhancement of carrier excitation efficiency near the metal-dielectric interface. We show that surface scattering is highly sensitive to the local field  polarization, and can be incorporated into metal dielectric function along with  phonon and impurity scattering.  We also show that the obtained surface scattering rate describes  surface-assisted plasmon decay (Landau damping) in nanostructures larger than the nonlocality scale. Our model can be used for calculations of plasmon-assisted hot carrier generation rates in photovoltaics and photochemistry applications.
\end{abstract}
\maketitle

\section{Introduction}

Plasmon-assisted hot carrier excitation and transfer across the interfaces \cite{nordlander-nn15} has recently attracted intense interest due to wide-ranging applications in  photovoltaics \cite{park-nl11,melosh-nl11,halas-science11,halas-nc13,lian-nl13,halas-nl13-2,clavero-np14,brongersma-nl14,atwater-nc14,halas-nc15} and photochemistry \cite{brongersma-nl11,moskovits-nl12,halas-nl13,moskovits-nn13,halas-jacs14}. In metal nanostructures with characteristic size $L$ below the diffraction limit,  i.e., $L<c/\omega$, where $c$ and $\omega$ are, respectively, the light speed and frequency, the light scattering is relatively weak, and the absorption is dominated by resonant excitation of surface plasmons, which subsequently decay into electron-hole (\textit{e-h}) pairs through several decay mechanisms depending on the size and shape of a plasmonic system \cite{halperin-rmp86,kresin-pr92,vallee-jpcb01,schatz-jpcb03,noguez-jpcc07}.  While in relatively large systems,  excitation of \textit{e-h} pairs with   optical energy $\hbar \omega$ is accompanied by momentum relaxation due to phonon and impurity scattering, for systems with characteristic size $L \sim 20$ or smaller, the dominant momentum relaxation channel is surface scattering that leads, in particular, to size and shape dependence of the plasmon linewidth \cite{klar-prl98,mulvaney-prl02,halas-prb02,klar-nl04,vallee-prl04,halas-nl04,hartland-pccp06,vallee-nl09,vanduyne-jpcc12,vallee-nl13,schatz-nl15}. Calculations of surface-assisted decay rate have been performed for some simple shapes (mostly spherical) within random phase approximation (RPA) \cite{kawabata-jpsj66,lushnikov-zp74,schatz-jcp83,barma-pcm89,yannouleas-ap92,eto-srl96,uskov-plasmonics13,khurgin-oe15,kirakosyan-prb16} or time-dependent local density approximation (TDDFT) \cite{jalabert-prb02,jalabert-prb05,yuan-ss08,vallee-jpcl10,lerme-jpcc11,li-njp13,nordlander-acsnano14} approaches, while for general shape systems, the classical scattering  model \cite{kreibig-zp75,ruppin-pss76,schatz-cpl83,schatz-jcp03,moroz-jpcc08} suggested the surface scattering rate in the form $\gamma_{cs}=A  v_{F}/L_{s}$, where $L_{s}$ is the ballistic electron scattering length, and the  constant $A$ includes the effects of surface potential, electron spillover, and dielectric environment \cite{kreibig-book}. Note, however, that  recent measurements of plasmon linewidth in nanostructures of various shapes revealed  significant discrepancies with a simple $L^{-1}$ dependence\cite{halas-nl04,vanduyne-jpcc12,vallee-nl13,schatz-nl15}.

Here we present a quantum-mechanical model for surface-assisted excitation of \textit{e-h} pairs by alternating local electric field of the form $\textbf{E}(\bm{r})e^{-i\omega t}$ created in the metal either by excitation of a plasmon or as a response to monochromatic external field. We show that surface contribution to the absorbed power due to energy transfer to the excited carriers is given by 
\begin{equation}
\label{power-surface}
Q_{s}=\frac{e^{2}}{2\pi^{2} \hbar} \frac{E_{F}^{2}}{ (\hbar\omega)^{2}} \! \int \!  dS  |E_{n}|^{2},
\end{equation}
where integration is taken over the metal surface, $E_{n}$ is the local field  component \textit{normal} to the interface, and $E_{F}$ is the Fermi energy in the metal. The above expression, which is derived within RPA approach, is valid for systems of arbitrary shape that are significantly larger than the nonlocality scale $v_{F}/\omega$ \cite{mortensen-pn13,mortensen-nc14}, i.e., for systems at least several nm large. We compute the surface enhancement of the carrier excitation efficiency for some common nanostructures and show that it is highly sensitive to the local field polarization and  system geometry.

\section{Theory}
We consider a metal nanostructure characterized by complex dielectric function $\varepsilon(\omega)=\varepsilon'(\omega)+i\varepsilon''(\omega)$ in a medium with dielectric constant $\varepsilon_{d}$, and, for simplicity,  restrict ourselves by  systems with a single metal-dielectric interface. The standard expression for absorbed power has the form \cite{landau}, 
\begin{equation}
\label{power-bulk}
Q=\frac{\omega\varepsilon''(\omega)}{8\pi} \! \int \! dV  |\textbf{E}|^{2}.
\end{equation}
where integration is carried over the metal volume, while the local field $\textbf{E}(\bm{r})$ is determined,  in the quasistatic limit,  by the Gauss's law $\bm{\nabla} [ \varepsilon' (\omega,\bm{r})\textbf{E}(\bm{r})]=0$ [here $\varepsilon (\omega,\bm{r})$ equals $\varepsilon(\omega)$ and $\varepsilon_{d}$  in the metal and dielectric regions, respectively]. The surface contribution to the absorbed power is $Q_{s}=\hbar\omega/\tau$, where $1/\tau$ is the first-order transition probability rate, leading to the standard RPA expression \cite{kirakosyan-prb16} 
\begin{equation}
\label{power-rpa}
Q_{s}=\pi \omega\sum_{\alpha\beta}|M_{\alpha\beta}|^{2}
\left [f(\epsilon_{\alpha})-f(\epsilon_{\beta})\right ]
\delta(\epsilon_{\alpha}-\epsilon_{\beta}+\hbar\omega).
\end{equation}
Here, $M_{\alpha\beta}=\int dV \psi_{\alpha}^{*}\Phi\psi_{\beta}$ is the matrix element of local potential $\Phi(\bm{r})$ defined as $e\textbf{E}=-\bm{\nabla} \Phi$ ($e$ is the electron charge),  $\psi_{\alpha}(\bm{r})$ and $\psi_{\beta}(\bm{r})$ are wave-functions for electron states with energies $\epsilon_{\alpha}$ and $\epsilon_{\beta}$, respectively, separated by $\hbar\omega$,  and $f(\epsilon)$ is the Fermi distribution function. For nanostructures of arbitrary shape, numerical evaluation of $M_{\alpha\beta}$ is a highly complicated task  due to the complexity of electron wave functions. However, as we show below, for the hard-wall confining potential, $Q_{s}$ can be derived in a closed form for any system significantly larger than $v_{F}/\omega$ (but still smaller than $c/\omega$).

Excitation of an \textit{e-h} pair with a large, compared to the electron level spacing, energy $\hbar\omega$ requires momentum transfer to the cavity boundary. We note that the boundary contribution  to $M_{\alpha\beta}$ can be presented as an integral over  the metal surface \cite{sm},
\begin{equation}
\label{matrix}
M_{\alpha\beta}^{s}= \frac{-e\hbar^{4}}{2m^{2} \epsilon_{\alpha\beta}^{2}} \! \int \! dS [\nabla_{n}\psi_{\alpha}(\bm{s})]^{*}E_{n}(\bm{s}) \nabla_{n}\psi_{\beta}(\bm{s}),
\end{equation}  
where $\nabla_{n}\psi_{ \alpha}(\bm{s})$ is the wave function derivative normal to the surface, $E_{n}(\bm{s})$  is the corresponding normal field component, $\epsilon_{\alpha\beta}=\epsilon_{\alpha}-\epsilon_{\beta}$ is the \textit{e-h} pair excitation energy, and $m$ is the electron mass.  Using this matrix element,  Eq.~(\ref{power-rpa})  takes the form
\begin{equation}
\label{Q-surface}
Q_{s}=\frac{e^{2}\hbar^{4}}{4\pi m^{4} \omega^{3}} \! \int \! \! \int \!  dS dS' E_{n}(\bm{s})E_{n'}^{\ast}(\bm{s}') F_{\omega}(\bm{s},\bm{s}'),
\end{equation}
where $F_{\omega}(\bm{s},\bm{s}')$ is the \textit{e-h} surface correlation function,
\begin{equation}
\label{F}
F_{\omega}(\bm{s},\bm{s}')=\! \int \! d\epsilon   f_{\omega}(\epsilon) \rho_{nn'}(\epsilon;\bm{s},\bm{s}')\rho_{n'n}(\epsilon+\hbar\omega;\bm{s}',\bm{s}),
\end{equation}
defined in terms of normal derivative of the electron cross density of states, $\rho(\epsilon;\bm{s},\bm{s}')=\text{Im} G(\epsilon;\bm{s},\bm{s}')$, at surface points: $\rho_{nn'}(\epsilon;\bm{s},\bm{s}')=\nabla_{n}\nabla'_{n'}\text{Im} G(\epsilon;\bm{s},\bm{s}')$. Here $G(\epsilon;\bm{s},\bm{s}')$ is the confined electron Green function, and the function $f_{\omega}(\epsilon)=  f(\epsilon)-f(\epsilon+\hbar\omega)$ restricts the initial energy of promoted electron to the interval $\hbar\omega$ below $E_{F}$.

To evaluate $Q_{s}$, we note that excitation of an \textit{e-h} pair with energy $\hbar\omega$ is accompanied by momentum transfer $\sim \hbar\omega/v_{F}$ and, hence, takes place in  a region of size $\sim v_{F}/\omega$, so that the \textit{e-h} correlation function $F_{\omega}(\bm{s},\bm{s}')$ peaks in the region $|\bm{s}-\bm{s}'|\lesssim v_{F}/\omega$ and rapidly oscillates outside of it (see below). At the same time, the local fields significantly change  on a much larger scale $\sim L$. Therefore,  for $L\gg v_{F}/\omega$,  the main contribution to the integral in Eq.~(\ref{Q-surface}) comes from the regions with $E_{n}(\bm{s})\approx E_{n}(\bm{s}')$,   so that $Q_{s}$ takes the form
\begin{equation}
\label{Q-surface1}
Q_{s}=\frac{e^{2}\hbar^{4}}{4\pi m^{4} \omega^{3}} \! \int \!  dS  |E_{n}(\bm{s})|^{2} \bar{F}_{\omega}(\bm{s}),
\end{equation}
where $\bar{F}_{\omega}(\bm{s})=\! \int \!  dS'   F_{\omega}(\bm{s},\bm{s}')$. Evaluation of $\bar{F}_{\omega}$ is based upon multiple-reflection expansion for the electron Green function $G(\epsilon;\bm{s},\bm{s}')$ in a hard-wall cavity \cite{balian-ap70}.  For system's characteristic size $L\gg \lambda_{F}$, where $\lambda_{F}$ is the Fermi wavelength, the main contribution comes from the direct and singly-reflected paths, while   the  higher-order reflections are suppressed as powers of $\lambda_{F}/L$. In the leading  order, we obtain $G(\epsilon;\bm{s},\bm{s}')=2G_{0}(\epsilon,\bm{s}-\bm{s}')$, where $G_{0}(\epsilon,r)=(m/2\pi \hbar^{2})e^{ik_{\epsilon}r}/r$, with $k_{\epsilon}=\sqrt{2m\epsilon}/\hbar$, is  the free electron Green function and factor 2 comes from equal contributions of the direct and singly-reflected paths at a surface point \cite{sm}.  It is now easy to see that the integrand of Eq.~(\ref{F}) peaks in the region  $|\bm{s}-\bm{s}'|\lesssim (k_{\epsilon+\hbar\omega}-k_{\epsilon})^{-1}$ and rapidly oscillates outside of it. For $\epsilon\sim E_{F}$ and $\hbar\omega/E_{F}\ll 1$, this sets the  scale $|\bm{s}-\bm{s}'|\sim v_{F}/\omega$  for the \textit{e-h} correlation  function $F_{\omega}(\bm{s},\bm{s}')$ in Eq.~(\ref{Q-surface}). Finally, for $L\gg v_{F}/\omega$, after computing normal derivatives in $\rho_{nn'}(\epsilon;\bm{s},\bm{s}')=2\nabla_{n}\nabla'_{n}\text{Im}G_{0}(\epsilon,\bm{s}-\bm{s}')$ relative to the tangent plane at a surface point \cite{sm}, we obtain $\bar{F}_{\omega}=(2m^{4}E_{F}^{2}/\pi \hbar^{8})\hbar\omega$, yielding Eq.~(\ref{power-surface}).

The surface contribution $Q_{s}$ to the full absorbed power $Q$  should be considered in conjunction with its bulk counterpart $Q_{0}$, i.e., $Q=Q_{0}+Q_{s}$. The bulk contribution is given by the standard expression (\ref{power-bulk}), where the metal dielectric function $\varepsilon(\omega)$ includes only the bulk processes. In the following, we adopt the Drude  dielectric function $\varepsilon(\omega)  = \varepsilon_{i}(\omega)-\omega_{p}^{2}/\omega(\omega+i\gamma)$, where $\varepsilon_{i}(\omega)$ describes interband transitions, $\omega_{p}$ is the plasma frequency, and $\gamma$ is the scattering rate. In fact, both bulk and surface contributions can be combined, in a natural way, within the general expression (\ref{power-bulk}). Indeed, using the relation $\omega_{p}^{2}=16e^{2}E_{F}^{2}/3\pi \hbar^{3}v_{F}$, the surface absorbed power (\ref{power-surface}) can be recast as
\begin{equation}
\label{Q-surface2}
Q_{s}
=\frac{3v_{F}}{32\pi}\frac{\omega_{p}^{2}}{\omega^{2}}\! \int \!  dS  |E_{n}|^{2}.
\end{equation}
Then, it is easy to see that the surface contribution can be incorporated into general expression (\ref{power-bulk}) for the absorbed power by modifying the Drude scattering rate as $\gamma=\gamma_{0}+\gamma_{s}$, where $\gamma_{0}$ is the usual bulk scattering rate and
\begin{equation}
\label{rate-ld}
\gamma_{s}=
\frac{3 v_{F}}{4}
 \frac{\int \! dS |E_{n}|^{2}}{\int \! dV |\textbf{E}|^{2}},
\end{equation}
is the surface scattering rate. Indeed, after this modification,  $Q_{s}$ can be  obtained from Eq.~(\ref{power-bulk})  as the first-order expansion term  in $\gamma_{s}$, indicating that the surface scattering mechanism should be treated on par with  phonon and impurity scattering. Note that  $\gamma_{s}$ is independent of the local field overall strength  but highly sensitive to its polarization relative to the metal-dielectric interface.

Turning to the surface-assisted  plasmon decay (Landau damping), the plasmon decay rate in any metal-dielectric structure is given by general expression \cite{shahbazyan-prl16} $\Gamma =  Q/U$, where $U$ is the plasmon energy \cite{landau},
\begin{align}
\label{energy-LL}
U 
=\frac{\omega}{16\pi} \frac{\partial \varepsilon'(\omega)}{\partial \omega}\! \int\! dV |\textbf{E} |^{2}.
\end{align}
Using Eq.~(\ref{power-bulk}) for the absorbed power,  the decay rate has the standard form,
\begin{equation}
\label{ld}
\Gamma = 2\varepsilon''(\omega)\left [\frac{\partial \varepsilon'(\omega)}{\partial \omega}\right ]^{-1}.
\end{equation}
Let us show that the full plasmon decay rate that includes both bulk and surface contributions is given by Eq.~(\ref{ld}), but  with  $\varepsilon(\omega)$ modified according to Eq.~(\ref{rate-ld}). Indeed, using Eq.~(\ref{Q-surface2}), the surface contribution to $\Gamma$ takes the form 
\begin{equation}
\label{ld1}
\Gamma_{s} =  \frac{Q_{s}}{U}=  \frac{2\omega_{p}^{2}\gamma_{s}}{\omega^{3}}\left [\frac{\partial \varepsilon'(\omega)}{\partial \omega}\right ]^{-1} ,
\end{equation}
where $\gamma_{s}$ is given by Eq.~(\ref{rate-ld}). The same expression is obtained, in the first order in $\gamma_{s}$, from Eq.~(\ref{ld}) with surface-modified $\varepsilon(\omega)$.  Note that for $\omega$ well below the interband transitions onset, the plasmon decay rate and  scattering rate coincide, $\Gamma\approx\gamma=\gamma_{0}+\gamma_{s}$.

\section{Applications}

Let us now discuss  the effect of field polarization and system geometry on the carrier excitation efficiency. The surface enhancement factor of the absorbed power is given by the ratio of full ($Q$) to bulk ($Q_{0}$) absorbed power, $M=Q/Q_{0}$, which, within RPA, takes a simple form
\begin{equation}
\label{enh}
M=1+\frac{\gamma_{s}}{\gamma_{0}}.
\end{equation}
Evaluation of surface scattering rate $\gamma_{s}$ is made more convenient by noting that the Gauss's law reduces the volume integral in Eq.~(\ref{rate-ld}) to the surface term, so that  
\begin{equation}
\label{rate-surface1}
\gamma_{s}=A\, v_{F} \, \frac{\int \! dS |\nabla_{n}\Phi |^{2}}{\int \! dS \Phi^{*} \nabla_{n}\Phi},
\end{equation}
where real part of the denominator is implied. This form  of $\gamma_{s}$ as the ratio of two surface integrals reflects the fact that  \textit{e-h} pairs are excited in a close proximity (within $v_{F}/\omega$) to the interface. The constant $A$, which equals $A=3/4$ for the hard-core confining potential, can be determined by matching Eq.~(\ref{rate-surface1}) with the corresponding rate for a spherical particle with radius $a$. A straightforward evaluation of Eq.~(\ref{rate-surface1}) recovers the standard result for a sphere $\gamma_{sp}=Av_{F}/a$, while in the recent TDLDA calculations for relatively large (up to $a=10$ nm)  spherical particles \cite{vallee-jpcl10,lerme-jpcc11}  the value $A\approx 0.32$ was obtained. Note that for systems, whose geometry permits separation of variables, the form (\ref{rate-surface1}) of $\gamma_{s}$ is especially useful since it leads to \textit{analytical} results  for some common structures that so far eluded attempts of any quantum-mechanical evaluation, as we illustrate below for metal nanorods and nanodisks.


%
\begin{figure}[tb]
\begin{center}
\includegraphics[width=0.9\columnwidth]{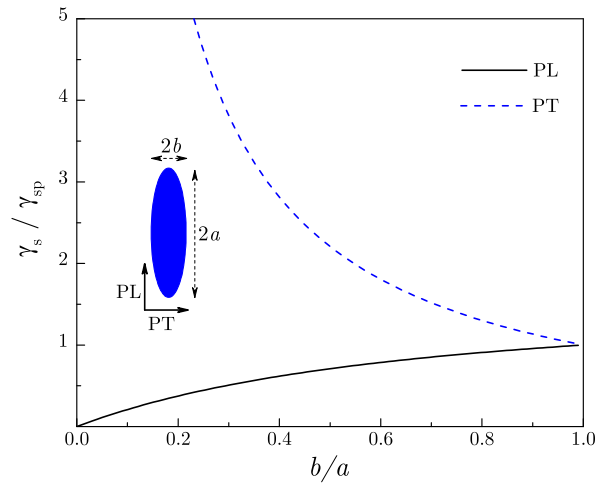}
\caption{\label{fig1}
Normalized rates for longitudinal and transverse dipole modes in prolate spheroidal particles (nanorods) relative to the spherical  particle rate are shown with changing aspect ratio $b/a$ Inset: Schematics for the mode polarizations. 
  }
\end{center}
\vspace{-6mm}
\end{figure}
%

%
\begin{figure}[tb]
\begin{center}
\includegraphics[width=0.9\columnwidth]{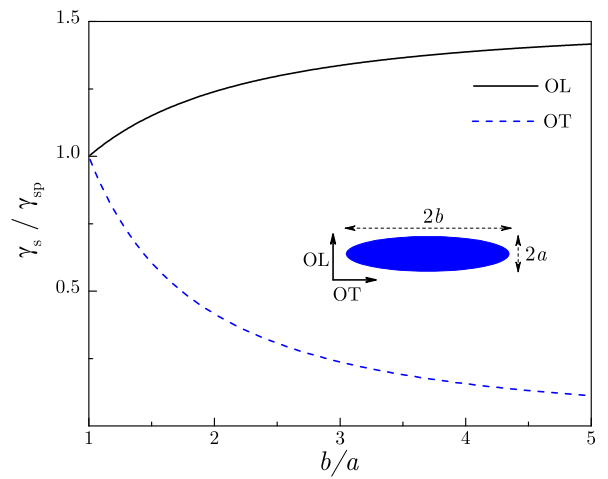}
\caption{\label{fig2}
Normalized rates for longitudinal and transverse dipole modes in oblate spheroidal particles (nanodisks) relative to the spherical  particle rate are shown with changing aspect ratio $b/a$ Inset: Schematics for the mode polarizations. 
  }
\end{center}
\vspace{-6mm}
\end{figure}
%

In Figs.~\ref{fig1} and \ref{fig2}, we show calculated surface scattering rates for nanorodes and nanodisks, which are modeled here by prolate and oblate spheroidal nanoparticles, respectively. These structures support longitudinal and transverse plasmon modes oscillating along the symmetry axis (semi-axis $a$) and within the symmetry plane (semi-axis $b$).  Using  Eq.~(\ref{rate-surface1}), $\gamma_{s}$ for all modes can be found in an analytical form \cite{sm}, but here only the results for the  dipole mode  are presented.  For a  nanorod  (prolate spheroid) with the aspect ratio $b/a<1$,  we obtain $\gamma_{s} =\gamma_{sp}f_{L,T}$, where 
\begin{align}
\label{rates-spheroid}
f_{L}
=\frac{3}{2\tan^{2}\!\alpha}\left [\frac{2\alpha}{\sin 2\alpha}-1
\right ],
~
f_{T}
=\frac{3}{4\sin^{2}\!\alpha}\left [1-\frac{2\alpha}{\tan 2\alpha}
\right ],
\end{align}
are the normalized rates for longitudinal and transverse modes  relative to the spherical particle rate, and $\alpha=\arccos (b/a)$ is the angular eccentricity. For a nanodisk (oblate spheroid) with    $b/a>1$, the normalized rates have the same form (\ref{rates-spheroid}) but with $\alpha=i\arccosh (b/a)$.  In Fig.~\ref{fig1}, we show the normalized rates in a prolate spheroid ($b/a<1$) for both longitudinal (PL) and transverse (PT) modes as the system shape evolves from a needle to a sphere. With changing aspect ratio $b/a$, the rates exhibit a dramatic difference in behavior depending  on the local field polarization.  As  nanorods become thinner,  the normalized rate  decreases for the PL mode but increases for the PT modes. These trends are reversed  for  nanodisks ($b/a>1$), shown in Fig.~\ref{fig2}: the normalized rates increase for the longitudinal (OL) mode and decrease for the transverse OT mode as the nanodisk radius increases (at fixed height).

The dominant factor that determines the surface enhancement of carrier excitation efficiency is the  local field polarization relative to the metal-dielectric interface. In a given system, the surface-assisted carrier excitation rate can be  manipulated in a wide range by choosing the electric field  orientation.   Note that recent measurements \cite{schatz-nl15} of plasmon spectra in cylinder-shaped nanorods and nanodisks revealed strong polarization dependence of   plasmon linewidth.

\section{Conclusions}

In summary, we developed a quantum-mechanical model for surface-assisted carrier excitation in plasmonic  nanostructures  of arbitrary shape. We derived explicit expressions  for surface absorbed power and scattering  rate that are highly sensitive to the local field polarization relative to the metal-dielectric interface. Our results can be used for calculations of  hot  carrier generation rates in photovoltaics and photochemistry  applications \cite{nordlander-nn15}. 



\acknowledgments

This work was supported in part by the National Science Foundation under grants  No. DMR-1610427 and No. HRD-1547754.


\end{document}